\documentclass[aps,prl,twocolumn,showpacs,preprintnumbers,amsmath,amssymb,superscriptaddress]{revtex4}%
\usepackage{epsfig}%
\usepackage{dcolumn}
\usepackage{bm}
\begin{document}%
\title{\Large\bf\boldmath 
Bound Magnetic Polarons in the $3d$-electron \\
Ferromagnetic Spinel Semiconductor 
CdCr$_2$Se$_4$ 
}%
\author{Vyacheslav G.~Storchak}
\email{mussr@triumf.ca}
\affiliation{Russian Research Centre ``Kurchatov Institute'',
 Kurchatov Sq.~1, Moscow 123182, Russia}
\author{Jess H.~Brewer}
\affiliation{Department of Physics and Astronomy,
 University of British Columbia,
 Vancouver, British Columbia, Canada V6T 1Z1}
\author{Peter L.~Russo}
\affiliation{TRIUMF, 4004 Wesbrook Mall, Vancouver, British Columbia, Canada V6T 2A3}
\author{Scott L.~Stubbs}
\affiliation{Department of Physics and Astronomy,
 University of British Columbia,
 Vancouver, British Columbia, Canada V6T 1Z1}
 \author{Oleg E.~Parfenov}
\affiliation{Russian Research Centre ``Kurchatov Institute'',
 Kurchatov Sq.~1, Moscow 123182, Russia}
\author{Roger L.~Lichti}
\affiliation{Department of Physics, Texas Tech University, 
 Lubbock, Texas 79409-1051, USA}
 \author{Tel'man G.~Aminov}
\affiliation{Institute for General and Inorganic Chemistry, 
 Moscow 119991, Russia}
\vfil 
\date{10 August 2009} 
\begin{abstract}
Muon spin rotation/relaxation spectroscopy 
has been employed to study electron localization 
around a donor center --- the positive muon --- 
in the $3d$ magnetic spinel semiconductor CdCr$_2$Se$_4$ 
at temperatures from 2 to 300~K in magnetic fields up to 7~T.  
A bound state of an electron around a positive muon 
--- a magnetic polaron --- 
is detected far above the ferromagnetic transition 
up to 300~K.  
Electron localization into 
a magnetic polaron occurs due to 
its strong exchange interaction 
with the magnetic $3d$ electrons 
of local Cr$^{3+}$ ions, 
which confines its wave function within $R \approx 0.3$~nm, 
allowing significant overlap with 
both the nearest and next nearest shells of Cr ions.  
\end{abstract}
\vfil
\pacs{ 75.50.Pp, 72.25.Dc, 72.20.Jv, 76.75.+i \\ }
%
%
\maketitle%
The semiconductors currently in use as working media 
in electronics and information technology 
(Si, Ge, GaAs {\it etc.}) are nonmagnetic; 
therefore the spin of the carriers 
has so far played a minor role in semiconductor devices.  
One way to enhance spin-related phenomena 
for semiconductor spintronics applications \cite{Prinz1998,Wolf2001}
is to incorporate magnetic ions (typically Mn) 
into nonmagnetic semiconductors 
to realize dilute magnetic semiconductors 
(DMS) \cite{Furdyna1988,Ohno1998}. 
The interplay between electric and magnetic properties 
in ferromagnetic (FM) Mn-doped  III-V DMS 
has recently been demonstrated 
\cite{Ohno1999,Tang2003,Chiba2003}.  
Combined with nonmagnetic semiconductors, 
these DMS may also serve as polarized spin injectors 
in spintronics devices \cite{Fiederling1999,Ohno1999a}.  
The FM in these $p$-type materials results from 
a long-range coupling between the Mn atoms 
mediated by holes generated by Mn substitution 
at the trivalent cation site 
\cite{Jungwirth2006}.  

Unfortunately, the ferromagnetism in 
III-V Mn-doped DMS 
is limited by low concentration of magnetic ions.  
Molecular beam epitaxy 
results in a non-equilibrium enhancement of the 
otherwise low solubility of transition metals in III-V hosts, 
but still allows incorporation of 
no more than about 7-8\% of Mn atoms; 
above this critical concentration 
Mn tends to cluster and phase separate \cite{Matsukura2002}.  
Even at lower concentrations, spatial homogeneity may 
be affected by adding Mn \cite{Jungwirth2006} 
and nanoscale-range magnetic inhomogeneities can occur 
\cite{Storchak2008}.  

By contrast, intrinsic magnetic semiconductors (MS) 
such as the $4f$ Eu chalcogenides 
or $3d$ Cr spinels exhibit spontaneous 
homogeneous ferromagnetic order without any doping.  
When doped, these MS show semiconducting behavior 
(both $n$- and $p$-type), 
which indicates strong mutual influence between 
electrical and magnetic properties \cite{Coey1999,Nagaev2002}.  
Successful demonstration of the epitaxial growth of 
EuO and CdCr$_2$Se$_4$ 
on technologically important semiconductors 
Si, GaN, GaAs and GaP \cite{Schmehl2007,Park2002,Kioseoglou2004} 
makes them very attractive working media for spintronics applications.  
These materials offer several important advantages over DMS, 
such as higher magnetization, spatial homogeneity 
and wider ranges of conductivity 
tunable by doping.  
The longer spin lifetimes and spin-scattering lengths 
of electrons in Si, GaAs and GaN 
\cite{Kioseoglou2004,Kikkawa1999,Oestreich1999}, 
as well as much higher electron mobilities 
compared with those of holes, 
makes the ability to support $n$-type conductivity in MS 
especially attractive for semiconductor spintronics.  

Since charge and spin transport for electrons in 
Si, GaAs, GaP and GaN is excellent, 
the effectiveness of prospective all-semiconductor 
spintronics devices is determined by 
the electron transport in doped magnetic semiconductors.  
These materials, however, 
can support 
states that lead to severe electron localization.  
In fact, MS provide optimal conditions for 
the formation of a new type of quasiparticle --- 
the {\sl magnetic polaron\/} (MP) --- 
in which conduction electron ``autolocalization'' 
stabilizes an atomic-scale region of the ferromagnetic phase 
well above $T_c$ \cite{deGennes1960,Kasuya1968,Nagaev2002}.  
This electron localization in MS profoundly modifies 
their magnetic, electrical and optical properties.  
In particular, such MP determine most of 
the transport properties of magnetic semiconductors, 
leading to metal-insulator transitions 
with a remarkable resistivity change of 
up to 13 orders of magnitude (in doped EuO) 
and colossal magnetoresistance which suppresses 
resistivity by 3-4 orders of magnitude 
in magnetic fields of $\sim 10$~T \cite{Nagaev2002}.  
Measurements of both resistivity and Hall effect 
in the ferromagnetic spinel CdCr$_2$Se$_4$ 
clearly show that these remarkable properties 
reflect changes in the density, not the mobility, 
of charge carriers \cite{Nagaev2002}.  
These effects can be explained 
in terms of electron localization into 
entities roughly the size of a unit cell: MP.  
In fact, the MP concept now forms the basis for 
numerous studies of MS and related materials \cite{vonMolnar2007}.  

A magnetic polaron is formed by an electron 
localized due to its strong exchange interaction $J$ 
with magnetic ions in its immediate environment, 
whose direct coupling is rather weak.  
Of relevance to the current study 
is the {\sl bound\/} MP,    
in which the increase in the kinetic energy of the electron 
due to localization is compensated by {\sl both\/} 
the $s$-$d(f)$ exchange interaction $J$ 
{\sl and\/} the Coulomb interaction 
with the corresponding donor, 
so that the net change in the energy 
\begin{equation}
 \Delta F = \frac{\hbar^2}{2m^* R^2} 
 - J \frac{a^3}{R^3} 
 - \frac{e^2}{\varepsilon R} 
\label{eq:FreeEnergy}
\end{equation}
has a minimum as a function of $R$ --- 
thus determining the radius of 
the electron confinement \cite{Nagaev2002,vonMolnar2007}.
The dominant exchange term is optimized by maximum overlap 
of the MP electron with nearby magnetic ions.  

Magnetic polarons have recently been detected in 
$4f$ magnetic semiconductors 
EuS, EuO, EuSe \cite{Storchak2009} and SmS \cite{Storchak2009a}, 
following muon spin rotation/relaxation ($\mu^+$SR) \cite{Brewer1994} 
experiments in nonmagnetic semiconductors 
\cite{Storchak1997,Storchak2003}, 
that revealed the details of electron capture 
to form a muonium (Mu $\equiv \mu^+ e^-$) atom.  
Formation and dissociation of Mu 
more generally models 
electron capture by and release from donor centers, 
since a positive muon acts in this respect just like 
any other Coulomb attractive 
impurity \cite{Storchak2004}.  
In non-magnetic semiconductors, formation of such Mu atoms 
is driven solely by the Coulomb interaction; 
in magnetic semiconductors, it is the combined effect 
of the Coulomb and exchange interactions 
which drives Mu formation \cite{Storchak2009,Storchak2009a} 

In treating the $s-d(f)$ exchange interaction, 
two limiting cases are important: 
when the $s$-electron bandwidth $W$ 
is large compared with $J$, 
and when $J \gg W$.  
The former case is typical for the $s-f$ exchange 
in rare-earth compounds where 
the extremely localized $f$-electrons 
are screened by electrons of other shells.  
In particular, in Eu compounds 
the $s$-electron delocalized in 
the rather wide (a few eV) hybridized $5d$-$6s$ band 
exchanges with the partially filled inner $4f$ shell, 
which is separated from the band states by 
completely filled $5s$ and $6p$ shells, 
thus reducing $J$ to a few tenths of an eV.  
The opposite case $W \ll J$ \cite{Anderson1955} 
provides the basis  for the well-known double exchange 
in transition metal compounds.
Perhaps surprisingly, 
this inequality is likely to be quite realistic 
in 
such compounds 
where the charge carriers are 
often of the same $d$-type as 
the localized spins of the magnetic ions.  
In particular, in CdCr$_2$Se$_4$ 
the Fermi level falls in the middle of 
the narrow $d$-band which lies 
just below a much broader unoccupied $s$-$p$ conduction band 
\cite{Continenza1994}.  

In this Letter, we present experimental evidence of 
severe electron localization into a magnetic polaron 
bound to 
a positive muon in the paramagnetic phase of 
the $3d$-electron magnetic semiconductor CdCr$_2$Se$_4$.  

CdCr$_2$Se$_4$ is a chalcogenide spinel with cubic symmetry 
(56 atoms per unit cell),  
a lattice constant of 10.72~\AA\ and a direct gap of about 1.5~eV.  
Its magnetic moment per chemical formula is close to 6~$\mu_B$, 
which corresponds to the sum of two Cr$^{3+}$ moments, 
each having a moment of 3~$\mu_B$.  
Ordering of the Cr$^{3+}$ moments into the FM phase 
occurs at $T_c = 130$~K.  
Above $T_c$ its paramagnetic susceptibility 
exhibits  Curie-Weiss behavior.  

Single crystals of CdCr$_2$Se$_4$ for the current study 
were grown by the closed-tube vapor transport technique.  
They all have perfect octahedral shape 
with typical sizes of 3-5~mm. 
They are slightly $n$-type with carrier concentration 
of about $10^{18}$~cm$^{-3}$ at room temperature. 
Magnetization (SQUID) measurements in $H=50$~Oe 
were used to determine $T_c = 130$~K for these crystals, 
in close agreement with literature data.  
Time-differential $\mu^+$SR experiments 
using 100\% spin-polarized positive muons 
were carried out on the M15 surface muon channel at TRIUMF 
using the {\it HiTime\/} $\mu^+$SR spectrometer.  

\begin{figure}[htb] 
\begin{center}
\vspace*{-22mm}
\hspace*{-10mm}
\includegraphics[width=1.35\columnwidth,angle=0]{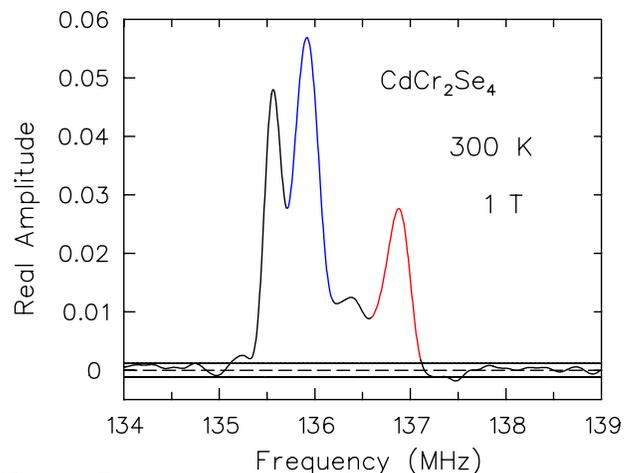}
\vspace*{-78mm}
\caption{
 Frequency spectrum of muon spin precession in CdCr$_2$Se$_4$ 
 in a transverse magnetic field of $H=1$~T at $T=300$~K.    
 The two-frequency      
 pattern 
 (blue and red online), 
 characteristic of the muon-electron bound state, 
 occurs at frequencies higher than 
 the narrower background signal.  
}
\end{center}
\vspace*{-6mm}
\label{fig:1-3lines}
\end{figure}

In a high magnetic field applied transverse 
to the initial muon polarization, 
the TF-$\mu^+$SR spectra exhibit two Mu-like signals 
shifted to higher frequencies relative 
to the narrow line positioned exactly at 
the diamagnetic frequency $\nu_\mu = \gamma_\mu B/2\pi$ 
(where $\gamma_\mu = 2\pi \times 135.5$~MHz/T 
is the muon gyromagnetic ratio 
and $B$ is the magnetic field) 
as shown in Fig.~1.  
This narrow line is a background signal 
from muons stopped outside the sample 
and provides a good reference 
since it does not depend on temperature.  
The two broader signals (blue and red online) 
at higher frequencies 
that depend on both temperature and magnetic field 
present a characteristic signature of 
the muon-electron bound state.  
For a $\mu^+ e^-$ (Mu) spin system
[as for $p^+ e^-$ (H)] 
governed by the Breit-Rabi Hamiltonian, 
these signals correspond to two muon spin-flip transitions 
between states with fixed electron spin orientation \cite{Brewer1994}.  

Accordingly, the rotating reference frame  
fits of the $\mu^+$SR spectra in the time domain 
at various temperatures 
show 3-frequency precession 
(1 background and 2 Mu-like signals, Fig.~2).  
The evolution of these signals is presented in the inset to Fig.~3. 
We claim that the two Mu-like lines are 
the spectroscopic signature of the magnetic polaron --- 
one electron localized around the positive muon 
by the combined effects of Coulomb and exchange interactions.  
In CdCr$_2$Se$_4$ this muon-bound MP forms at temperatures 
well above $T_c$, up to at least 300~K.  
The shift of the centroid of the two-line MP spectrum, 
the splitting and linewidths all scale with 
the bulk magnetic susceptibility over the 160-300 K temperature range, 
implying a common origin --- magnetic polarons 
and their orientational dynamics.

\begin{figure}[t]   
\begin{center}
\includegraphics[width=0.75\columnwidth,angle=0]{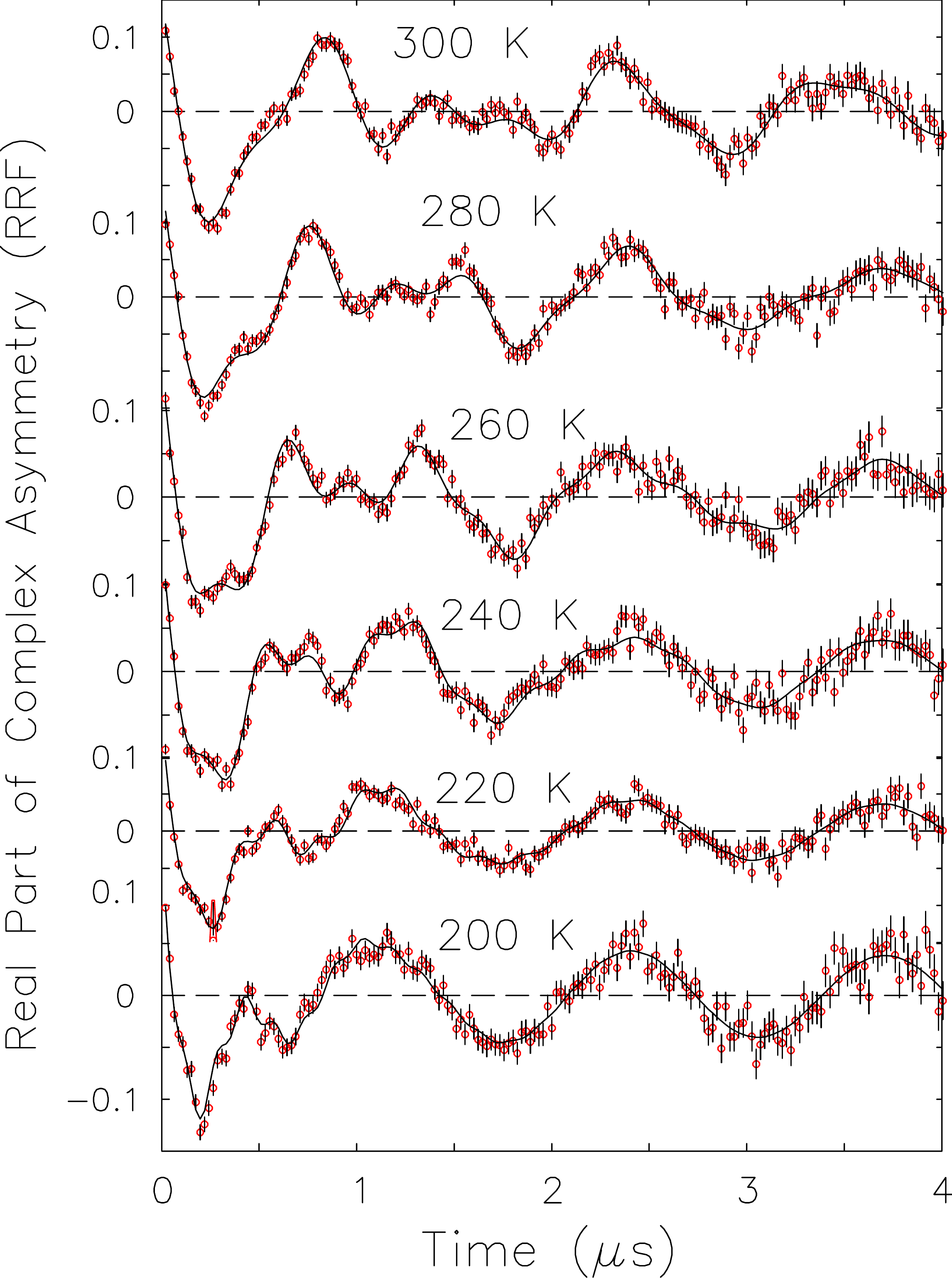}
 \vspace*{-2mm}
\caption{
 Time-domain fits of TF-$\mu$SR spectra in the rotating reference frame 
 in a field of $H=1$~T at different temperatures.  
}
\end{center}
\vspace*{-10mm}
\label{fig:2}
\end{figure}

Similar 2-frequency signals 
originating from 
magnetic polarons 
are detected in the $4f$ magnetic semiconductors 
EuS, EuO, EuSe \cite{Storchak2009} and SmS \cite{Storchak2009a}.  
In all MS studied, the MP lines are shifted with respect 
to the reference signal, 
reflecting the local ferromagnetic environment around the muon. 
In Eu chalcogenides, 
the MP lines exhibit negative shifts 
with respect to the reference signal, 
indicating that the MP electron spin is opposite to 
the net polaron spin.  
Addition of the MP electron effectively reduces the spin of 
the neighboring Eu ion from 7/2 
(according to Hund's rule, 
the maximum allowed spin is for 
a half-filled $f$-shell) 
to 3 
due to Pauli exclusion.  
By contrast, in Cr spinels the MP electron 
is bound to have its spin parallel to the net MP spin 
(Hund's rule for a less than half-filled $d$-shell), 
which effectively increases the spin of 
one of the neighboring Cr ions from its original value of 3/2 
to 2.  
Accordingly, in CdCr$_2$Se$_4$   
the FM shift is positive (see Fig.~1).  
The absolute value of this shift 
at room temperature and 1~T is about 0.007~T.  

The splitting $\Delta\nu$ between the two MP lines 
provides information on the muon-electron 
hyperfine coupling $A$, 
which is determined by the probability density of the 
electron wavefunction at the muon \cite{Brewer1994}.  
Figure~3 shows the temperature dependence of this splitting.  
A distinctive feature of magnetic semiconductors in general, 
and CdCr$_2$Se$_4$ in particular, is a strong dependence 
of the conduction electron energy on the magnetization 
due to the exchange interaction between the mobile electron 
and localized $d(f)$ spins, 
the minimum electron energy being achieved 
at the ferromagnetic ordering \cite{Nagaev2002}.  
In the paramagnetic state, an ``extra'' electron tends to 
establish and support this ordering, 
thus forming a FM ``droplet'' (MP) 
over the extent of its wave function.  
The exchange contribution to the localization 
[see Eq.~(\ref{eq:FreeEnergy})] 
amounts to the difference between 
the paramagnetic disorder of the CdCr$_2$Se$_4$ 
and the enhanced (FM) order in the MP.  
By contrast, in the FM state 
the exchange contribution to the localization is negligible, 
as the lattice spins are already aligned.  
In fact, as the magnetization develops towards low temperature, 
the exchange contribution to electron localization diminishes 
and therefore can no longer compensate 
the increase of its kinetic energy.  
The electron thus avoids strong localization 
as the temperature approaches $T_c$ from above.  
Accordingly, we do not detect the MP lines 
below about 150~K  
(see Fig.~3).  

\begin{figure} [t]  
\begin{center}
\vspace*{-8mm}
\hspace*{-11mm}
\includegraphics[width=0.86\columnwidth,angle=-90]{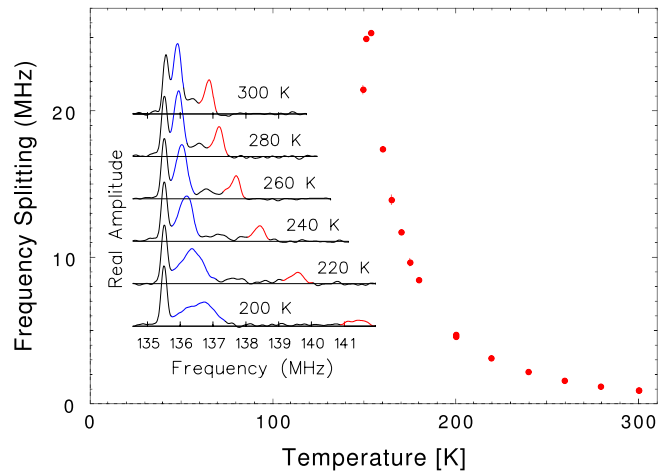}
\vspace*{-10mm}
\caption{
 Temperature dependence of the splitting between MP lines 
 in a magnetic field of $H=1$~T.  
 The inset shows the evolution of 
 the frequency spectra with temperature.  
}
\end{center}
\vspace*{-10mm}
\label{fig:3}
\end{figure}

Temperature and magnetic field dependences 
of the MP signal frequency splitting $\Delta\nu$ 
provide information on the characteristic size 
(the localization radius for electron confinement) 
of the MP in CdCr$_2$Se$_4$.  
%
Since the Mu electron spin 
is ``locked'' to the MP moment 
and the experiment samples the component along the applied field, 
the observed splitting scales with the mean   
orientation of the MP moment with respect to the external field, 
reaching the intrinsic Mu hyperfine constant only 
for complete alignment 
(see Appendix of Ref.\cite{Storchak2009}). 
Amplitudes for the two-frequency spectrum qualitatively scale 
as expected for a bound electron whose spin is locked to that of the 
polaron core, providing strong support for this picture.     
Within a mean field approximation, $\Delta\nu$ is thus 
proportional to a Brillouin function \cite{Smart1966,Storchak2009a} .  
For $g \, \mu_{_{\rm B}} B \ll k_{_{\rm B}} T$, 
this    
is linear in $B/(T-\Theta)$ (see Fig.~4): 
\begin{equation}
 \Delta\nu = A \left[
 \frac{g \, \mu_{_{\rm B}} B}{3k_{_{\rm B}} (T-\Theta)} 
 \right] ({\cal S}+1) \; ,
\label{eq:Splitting}
\end{equation}
where $\Theta=140$~K is the paramagnetic Curie temperature of 
CdCr$_2$Se$_4$ and ${\cal S}$ is the net polaron spin. 
At low $T - \Theta$ and high $B$, 
$\Delta\nu$ 
saturates at a value of $A$ \cite{Brewer1994,Storchak2009,Storchak2009a}.  

The vacuum state of a Mu atom is characterized by 
its hyperfine coupling $A_{\rm vac} = 4463$~MHz 
which corresponds to electron confinement within 
$R_{\rm Bohr} = 0.0529$~nm.  
In a solid nonmagnetic medium, Mu usually has $A < A_{\rm vac}$ 
so that, in a magnetic field high enough to satisfy 
$\gamma_\mu B/2\pi \gg A$, 
the splitting is $\Delta \nu = A$, 
independent of temperature and magnetic field 
\cite{Brewer1994,Storchak2009,Storchak2009a}.  
Figure 4 shows the MP frequency splitting in CdCr$_2$Se$_4$ 
as a function of both $1/(T-\Theta$) and $H$.  
Saturation is clearly seen in both the temperature dependence 
and the magnetic field dependence at $T=200$~K: 
on both plots $\Delta\nu$ levels off to give 
$A = 24{\pm}2$~MHz.  

\begin{figure}[t] 
\begin{center}
\vspace*{-6mm}
\hspace*{-11mm}
\includegraphics[width=0.86\columnwidth,angle=-90]{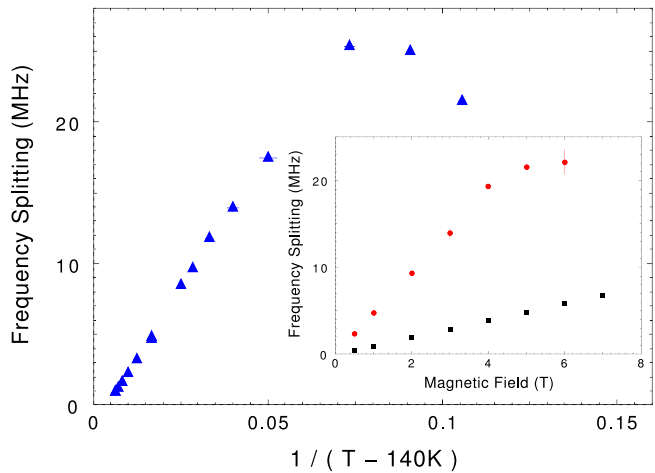}
\vspace*{-9mm}
\caption{
 Dependence of the splitting between MP lines on inverse temperature 
 in a magnetic field of $H=1$~T.  
 Inset: magnetic field dependence of the same splitting 
 at $T=200$~K (circles) and $T=300$~K (squares).  
}
\end{center}
\vspace*{-10mm}
\label{fig:4}
\end{figure}

This value of $A$ gives a measure of the electron confinement
around the muon at T=200~K: 
assuming an expanded hydrogen-like MP  wavefunction, 
the hyperfine coupling $A$ scales as $1/R^3$, 
where $R$ is the characteristic radius 
of the corresponding $1s$ electron wave function. 
We find $R \approx 0.3$~nm, which falls between 
the nearest neighbor (NN) and next nearest neighbor (NNN) 
ion coordination spheres.   
This value of $R$ is about an order of magnitude less than 
that calculated in \cite{Yanase1972} for temperature just above T$_c$.
It is known, however, that $R$ grows very fast as one approaches T$_c$ from
above \cite{Nagaev2002,Kasuya1968}.  
The value of the MP spin extracted from the slopes of 
both linear dependences of $\Delta\nu$ on $T$ and $H$
at higher temperature and lower magnetic field 
(see Fig.~4 and the inset for $T=200$~K)
using Eq.~(\ref{eq:Splitting}) amounts to ${\cal S} = 30 \pm 4$.  
This is reasonably consistent with a fully polarized core 
of 4 NN and 12 NNN Cr$^{3+}$ ions, 
yielding a net spin of 24, plus a partially ordered halo.
 
Here we find an important difference between MP in $4f$ and $3d$ MS: 
while in $4f$ MS the radius of the electron confinement 
is strongly ``glued'' to the corresponding NN 
coordination sphere \cite{Storchak2009a,Storchak2009}
because of the requirement of maximum electron overlap with
extremely localized ($\sim 0.03$~nm) $4f$ ion electrons, 
in $3d$ MS this requirement is much less stringent because of 
the significantly larger ``spread'' of the $3d$ wavefunction.  
In CdCr$_2$Se$_4$, therefore, $R$ falls in between 
the NN and NNN  Cr$^{3+}$ ions.  
More generally, exchange-driven electron 
localization in $3d$ magnets 
might enhance FM coupling between host magnetic ions 
by formation of MP around impurity atoms.


\vspace{-0.5mm}

This work was supported by
the Natural Sciences and Engineering Research Council of Canada 
and the U.S. Department of Energy (Grant DE-SC0001769).  



\begin{thebibliography}{99}

\bibitem{Prinz1998}
G.A.~Prinz,
{\sl Science} {\bf 282}, 1660  (1998).

\bibitem{Wolf2001}
S.A.~Wolf {\it et al.},
{\sl Science} {\bf 294}, 1488 (2001).

\bibitem{Furdyna1988}
J.K.~Furdyna,
{\sl J. Appl. Phys.} {\bf 264}, R29  (1988).

\bibitem{Ohno1998}
H.~Ohno,
{\sl Science} {\bf 281}, 951  (1998).

\bibitem{Ohno1999}
H.~Ohno {\it et al.},
{\sl Nature} {\bf 408}, 944 (1999).

\bibitem{Tang2003}
H.X.~Tang {\it et al.},
{\sl Phys.Rev.Lett.} {\bf 90}, 107201 (2003).

\bibitem{Chiba2003}
D.~Chiba {\it et al.},
{\sl Science} {\bf 301}, 943 (2003).

\bibitem{Fiederling1999}
 R.~Fiederling {\it et al.,}
 {\sf Nature} {\bf 402}, 787 (1999).

\bibitem{Ohno1999a}
 Y.~Ohno {\it et al.,}
 {\sf Nature} {\bf 402}, 790 (1999).

\bibitem{Jungwirth2006}
 T.~Jungwirth {\it et al.,}
  {\sf Rev. Mod. Phys.} {\bf 78}, 809 (2006).

\bibitem{Matsukura2002}
F.~Matsukura, H.~Ohno and T.~Dietl,
in {\sl Handbook of Magnetic Materials},
(Ed. K.H.J.~Buschow), 2002. 

\bibitem{Storchak2008}
 V.G.~Storchak {\it et al.,}
 {\sf Phys. Rev. Lett.} {\bf 101}, 027202 (2008).

\bibitem{Coey1999} 
 J.M.D.~Coey {\it et al.,}
 {\sf Adv. Phys.} {\bf 48}, 167 (1999).  

\bibitem{Nagaev2002} 
 E.L. Nagaev, 
 in {\sf Magnetic Semiconductors} (London: Imperial College Press, 2002). 

\bibitem{Schmehl2007}
 A.~Schmehl {\it et al.,}
 {\sf Nat. Mater.} {\bf 6}, 882 (2007).
 
\bibitem{Park2002}
 Y.D.~Park {\it et al.,}
 {\sf Appl. Phys. Lett.} {\bf 81}, 1471 (2002).
 
\bibitem{Kioseoglou2004}
 G.~Kioseoglou  {\it et al.,}
 {\sf Nat. Mater.} {\bf 3}, 799 (2004).

\bibitem{Kikkawa1999}
 J.~Kikkawa and D.~Awschalom,
 {\sf Nature} {\bf 397}, 139 (1999).

\bibitem{Oestreich1999}
 M.~Oestreich,
 {\sf Nature} {\bf 402}, 735 (1999).

\bibitem{deGennes1960}
 P.G.~de Gennes,
 {\sf Phys. Rev.} {\bf 118}, 141 (1960).

\bibitem{Kasuya1968} 
 T.Kasuya and A.Yanase, 
 {\sf Rev. Mod. Phys.} {\bf 40}, 684 (1968). 


\bibitem{vonMolnar2007} 
 S. von Moln{\'a}r and  P.A. Stampe, 
in {\sf Handbook of Magnetism and Advanced Magnetic
 Materials}, eds.  H.~Kronmueller and S.~Parkin 
 (John Wiley \& Sons, Ltd., 2007).


\bibitem{Storchak2009} 
 V.G.~Storchak {\it et al.,}
{\sf Phys.~Rev.~B} {\bf 80}, 235203 (2009); 
{\sf Phys. B} {\bf 404}, 899 (2009).

\bibitem{Storchak2009a} 
V.G.~Storchak {\it et al.,}
      {\sl Phys. Rev.} {\bf B79}, 193205 (2009).
 
\bibitem{Brewer1994} 
 J.H. Brewer, 
 {\sf Encyclopedia of Applied Physics} {\bf 11}, 23-53 
  (VCH Publishers, New York, 1994).  

\bibitem{Storchak1997} 
 V.G. Storchak,  {\it et al.}, 
 {\sf Phys. Rev. Lett.} {\bf 78}, 2835 (1997).  


\bibitem{Storchak2003} 
 V.G. Storchak {\it et al.}, 
 {\sf Phys. Rev. B} {\bf 67}, 121201 (2003).  

\bibitem{Storchak2004} 
 V.G. Storchak {\it et al.}, 
 {\sf J. Phys.}, {\bf 16}, S4761 (2004).  

\bibitem{Anderson1955}
 P.W.~Anderson and H.~Hasegawa,
 {\sf Phys. Rev.} {\bf 100}, 675 (1955).  

\bibitem{Continenza1994}
 A.~Continenza {\it et al.},
  {\sf Phys. Rev. B} {\bf 49}, 2503 (1994).  

\bibitem{Smart1966} 
 J.S.~Smart, 
 {\it Effective Field Theories of Magnetism}, 
 (Saunders, Philadelphia, 1966).  

\bibitem{Yanase1972}
 A.~Yanase,
 {\sf Int. J. Magn.} {\bf 2}, 99 (1972).

\end{thebibliography}
\end{document}